\documentclass[letter]{aa}

\usepackage{appendix}
\usepackage[english]{babel}
\usepackage[babel=true]{csquotes}

\usepackage[varg]{txfonts}
\usepackage{amsmath}
\usepackage{geometry} 
\usepackage{graphicx} 
\usepackage{fancyhdr}
\usepackage[amssymb]{SIunits}
\usepackage{amsfonts}
\usepackage[version=3]{mhchem}
\usepackage{sectsty}
\usepackage{numprint}
\usepackage{wasysym}
\usepackage{multirow}
\usepackage{fancyhdr}
\usepackage{placeins}

\usepackage[varg]{txfonts}
\usepackage{natbib}
\bibpunct{(}{)}{;}{a}{}{,} % to follow the A&A style

\makeatother

\begin{document}

\title{Microlensing of the broad-line region in the quadruply imaged quasar HE0435-1223\thanks{Based on observations made with the ESO-VLT, Paranal, Chile; Proposal 084.B-0013 (PI: Rix).}}

\author{ L.~Braibant\inst{1}, D.~Hutsemékers\inst{1}, D. Sluse\inst{2}, T. Anguita\inst{3}, C. J. Garc\'{i}a-Vergara\inst{4}}
\authorrunning{Braibant et al.}

\institute{
\inst{1} F.R.S.-FNRS, Institut d’Astrophysique, Universit\'{e} de Li\`{e}ge, All\'{e}e du 6 Août 17, B5c, 4000 Li\`{e}ge, Belgium \\
\inst{2} Argelander-Institut für Astronomie, Auf dem Hügel 71, 53121 Bonn, Germany \\
\inst{3} Departamento de Ciencias Fisicas, Universidad Andres Bello, Av. Republica 252, Santiago, Chile \\
\inst{4} Instituto de Astrof\'{i}sica Pontificia Universidad Cat\'{o}lica de Chile, Avenida Vicu\~{n}a Mackenna 4860, Santiago, Chile
}

\abstract{Using infrared spectra of the z=1.693 quadruply lensed quasar HE0435-1223 acquired in 2009 with the spectrograph SINFONI at the ESO Very Large Telescope, we have detected a clear microlensing effect in images $A$ and $D$. While microlensing affects the blue and red wings of the \ce{H\alpha} line profile in image $D$ very differently, it de-magnifies the line core in image $A$. The combination of these different effects sets constraints on the line-emitting region; these constraints suggest that a rotating ring is at the origin of the \ce{H\alpha} line. Visible spectra obtained in 2004 and 2012 indicate that the \ce{MgII} line profile is microlensed in the same way as the \ce{H\alpha} line. Our results therefore favour flattened geometries for the low-ionization line-emitting region, for example, a Keplerian disk. Biconical models cannot be ruled out but require more fine-tuning. Flux ratios between the different images are also derived and confirm flux anomalies with respect to estimates from lens models with smooth mass distributions.}
\maketitle

\section{Introduction}

Gravitational microlensing by compact objects in lensing galaxies is a powerful tool for probing the inner regions of distant quasars. Microlenses typically magnify regions of the source on scales of a few tens of light days (e.g., \citealp{2010SchmidtWambganss}). The region that generates the optical and UV continuum (i.e., the accretion disk) is therefore very likely (de-)magnified by microlensing, which can be observed as a (de-)magnification of the continuum spectrum since micro-images are not resolved. The broad-line region (BLR) can also be affected by microlensing. \citet{2012Sluseal} have shown that microlensing of the BLR is commonly detected in systems exhibiting microlensing of the continuum. Interestingly, microlensing of the BLR can significantly alter the broad emission-line profiles (e.g., \citealp{2002Abajasal}, \citealp{2004LewisIbata}) so that it may be possible to exploit these spectral differences to probe the BLR structure.

We study the broad emission lines of the gravitational lens system HE0435-1223 discovered by \citet{2001Wisotzkial}. It is composed of four quasar images of similar brightness, separated by about $\unit{1.5}{\arcsec}$ and arranged with quasi-perfect symmetry around the elliptical lensing galaxy. The quasar is at redshift $z_s = 1.693$ \citep{2012Sluseal}. The redshift of the lens is $z_l = 0.4546$ \citep{2005Morganal}. HE0435-1223 appears to be a good candidate for studying microlensing effect on the line profiles (e.g., \citealp{2003Wisotzkial}, and references therein).

In the following, we mainly focus on the distortions of the \ce{H\alpha} line profile in the spectra of the quasar-lensed images observed between October and December 2009. We also examine \ce{MgII} line profiles observed in November and December 2004, and in September 2012. We finally draw some conclusions about the geometry and kinematics of the BLR.

\section{Data collection and processing}
\label{sec:data}

\subsection{Infrared spectra}

We used archive Integral Field Spectroscopy data of HE0435-1223 secured between Oct. 19 and Dec. 15, 2009 at the Very Large Telescope (Table~\ref{tab:obsHE0435}). The system has been observed with SINFONI, using the $\unit{3}{\arcsec} \times \unit{3}{\arcsec}$ field-of-view (FOV) with $\unit{0.1}{\arcsec}$ spatial resolution. The H-band grism leads to a spectral coverage of $\unit{[1.45,1.85]}{\micro\metre}$, which contains the broad \ce{H\alpha} emission line. Due to offsets between exposures, some lensed images were occasionally outside the FOV. Only the spectra of the quasar images entirely enclosed in the FOV were extracted.

The data were calibrated and 3D data cubes, in which each plane is a monochromatic FOV, were built with the ESORex SINFONI pipeline (version 2.2.9). Cosmic rays were removed from every monochromatic FOV using the \texttt{la\_cosmic} procedure \citep{2001Vandok_lacosmic}.  In addition, the SINFONI FOV is not uniformly illuminated. Since this is not corrected for by the pipeline, we applied an empirical illumination correction: we divided each monochromatic FOV by a normalized median FOV of the infrared empty sky. Nevertheless, as a precautionary measure, we discarded the spectra of the images whose peak is located on slitlets $\#$8, $\#$9, and $\#$10, which show an illumination minimum, and on slitlet $\#$25, which occasionally may show a defect. %when occasionally showing a defect.

The spectra of the quasar images were extracted from the data cubes by fitting a simplified model of the system to each monochromatic FOV, using a modified \texttt{MPFIT} package \citep{2009Markwardt}. In this model, the quasar images were modelled with identical 2D Gaussian point spread functions (PSFs), whose relative positions were fixed to the astrometry of \citet{2011Courbinal}. The lensing galaxy was modelled with a de Vaucouleurs profile with $r_{eff} = \unit{1.5}{\arcsec}$ \citep{1948DeVaucouleurs}, convolved with the PSF. The fitting procedure was iterative: the values of the model parameters determined at a given iteration were used as starting values for the following fit after they were smoothed over the wavelength with a median filter and fitted with a spline function. The PSF widths and the position of the whole system were forced to vary between physical boundaries that were shrunk at every iteration. 

\begin{table}
\begin{footnotesize}
\centering
\caption{Flux ratios measured in the continuum.}
\begin{tabular}[htpb]{l c c c c c c}
\hline \\ [-2.0ex]
Date & Spectra & A/B\tablefootmark{a} & C/B\tablefootmark{b} & D/B\tablefootmark{c} & C/A\tablefootmark{d} & D/A\tablefootmark{e} \\
\hline  \\ [-2.0ex]
19/10 & A, B, C, D & $1.424$ & $1.047$ & $0.762$ & $0.736$ & $0.535$ \\
19/10 & A, C & - & - & - & $0.658$ & -  \\
19/10 & A & - & - & - & - & - \\
19/10 & A, B, D & $1.356$ & -  & $0.937$ & -  & $0.691$ \\
09/12 & A, C, D & -  & -  & -  & $0.823$ & $0.555$ \\
09/12 & A, C &  - & -  & -  & $0.745$ & -  \\
09/12 & A & - & - & - & - & - \\
09/12 & A, B, D & $1.354$ & -  & $0.837$ & -  & $0.618$ \\
09/12 & A, C, D & -  & -  & -  & $0.820$ & $0.572$ \\
09/12 & A, C &  - & -  & -  & $0.789$ & -  \\
09/12 & A & - & - & - & - & - \\
09/12 & A, B, D & $1.311$ & -  & $0.720$ & -  & $0.549$ \\
10/12 & B & - & - & - & - & - \\
10/12 & B, C & -  & $1.045$ & -  & -  & -  \\
15/12 & A, B, C, D & $1.353$ & $0.999$ & $0.720$ & $0.738$ & $0.532$ \\
15/12 & A, C & -  & -  & -  & $0.763$ & -  \\
15/12 & A & - & - & - & - & - \\
15/12 & A, B, D & $1.376$  & -  & $0.778$ & -  & $0.565$ \\
\hline
\end{tabular}
\tablefoot{
We list for each science exposure acquired with SINFONI in 2009 the observing date (col. 1), the images with extracted spectra (col. 2), and the flux ratios measured in the continuum (col. 3-7).\\
Error bars : \tablefoottext{a}{$\pm 0.14$}
\tablefoottext{b}{$\pm 0.11$}
\tablefoottext{c}{$\pm 0.08$}
\tablefoottext{d}{$\pm 0.08$}
\tablefoottext{e}{$\pm 0.06$}
}
\label{tab:spec_ratios}
\end{footnotesize}
\end{table}

\begin{figure}
\includegraphics[width=0.5\textwidth, height=7.5cm]{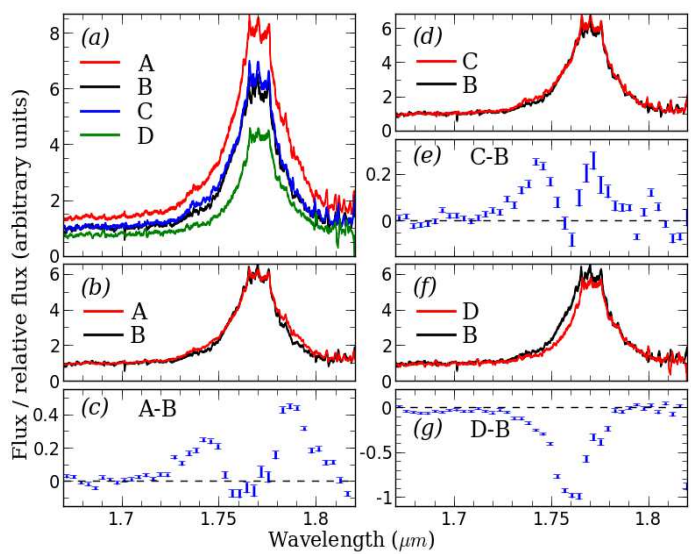}
\caption{\textit{(a)} Median spectra of the QSO-lensed images. \textit{(b)} The spectra of images $A$ and $B$ normalized so that their continua have a unit average intensity and are superimposed, and \textit{(c)} the spectral differences between the \ce{H\alpha} line profiles in the normalized spectra of images $A$ and $B$, plotted with error bars for a wavelength bin of $\unit{0.004}{\micro\meter}$. \textit{(d)} and \textit{(e)} Same for the pair of images B and C. \textit{(f)} and \textit{(g)} Same for the pair of images B and D. Spectra have been filtered with a five-pixel-wide median filter for clarity. %The thickness of the curves represents the $1\sigma$-error bars.
}
\label{fig:he0435_median}
\end{figure}

Table~\ref{tab:spec_ratios} provides for each science exposure the list of usable spectra and the measured image flux ratios. These ratios were obtained by fitting the previous model to a \enquote{pure continuum} FOV, derived by co-adding all the monochromatic FOVs whose wavelengths are located in the range $\unit{[1.523,1.664]}{\micro\metre}$. The errors on flux ratios were estimated by multiplying the flux ratios with the dispersion of the $F_A/F_A$ ratio ($\sigma_{A/A}=0.106$), computed for all the spectra of image~$A$ secured on Dec. 9, 2009\footnote{The night with the most observed spectra.}.

The spectra were normalized so that their continua have a unit average intensity. Since the normalized spectra of each quasar image are compatible within $3\sigma$-error regardless of the observing date, we increased the signal-to-noise ratio by computing a median normalized spectrum over all the epochs for each lensed image. The median spectra of images $A$, $C$, and $D$ were then multiplied by the mean flux ratio in the continuum, $A/B =1.36 \pm 0.14$, $C/B=1.03 \pm 0.11$, and $D/B=0.79 \pm 0.08$. The renormalized median spectra are plotted in Fig.~\ref{fig:he0435_median}$a$.

%{$\bullet$ \raggedleft{}\textbf{Visible spectra}}
\subsection{Visible spectra}

We also examined visible spectra of $B$ and $D$ obtained in 2004 with the ESO-VLT FORS1 instrument (Fig.\ref{fig:comp_Dref_vis_ir}$a$). The observations are summarized in Table~\ref{tab:obsHE0435}. More details about the data reduction can be found in \citet{2006Eigenbrod} and \citet{2012Sluseal}.

Additional spectroscopic observations of HE0435-1223 were acquired in 2012 using the IMACS spectrograph (long-camera mode) at Magellan-Baade (see Table~\ref{tab:obsHE0435}). The slit was oriented through components $A$-$D$ on one hand and through components $B$-$C$ on the other hand.

The Magellan data were reduced using the \textsc{COSMOS}\footnote{http://code.obs.carnegiescience.edu/cosmos} package. The 2D slit spectra were extracted in two steps. First, for each wavelength bin, we fitted a sum of two identical Moffat profiles on the two quasar images, with no prior on the image positions. The image centers and Moffat widths, derived after that first step, were smoothed with a median filter and fitted with a spline function to set the initial conditions of the second fit. In this second fit, the parameters were constrained to agree with the smoothed values within 5\%. The spectra of images $D$ and $A$ are presented in Fig.~\ref{fig:comp_Dref_vis_ir}$c$.

\section{Microlensing in HE0435-1223}
\label{sec:microlensing_decomposition}

\subsection{Evidence for microlensing}

Fig.~\ref{fig:he0435_median}$b-g$ highlight the differences between the \ce{H\alpha} line profiles observed in the median spectra of the different quasar images, normalized in the continuum. Following \citet{2011Courbinal}, who identified image $B$ as the \enquote{least affected by stellar microlensing}, we assumed that the spectrum of image $B$ is not altered by microlensing and used it as the reference to which we compare the spectra of the other quasar images.

The spectral-line profiles of $A$ and $D$ differ significantly from the profiles of the two other images, $A$ having a brighter red wing (see Fig.~\ref{fig:he0435_median}$b,c$) and $D$ a fainter blue wing (see Fig.~\ref{fig:he0435_median}$f,g$). We computed median spectra of the quasar images for each individual observing night. At every epoch, image $A$ displays a prominent red \ce{H\alpha} wing and image $D$ a dimmer blue \ce{H\alpha} wing than image $B$; these differences are therefore probably not caused by the intrinsic variability of the quasar. This interpretation is supported by R-band light curves published in \citet{2011Courbinal}, which show that the fluxes of the images have varied by less than 10\% during the period of the SINFONI observations.

The line profiles of images $B$ and $C$ only show small differences in their blue wing (see Fig.~\ref{fig:he0435_median}$d,e$). The amplitude of the differences between $B$ and $C$ varies from one observing date to another, which indicates that they may be (partly) caused by intrinsic variability. 
Comparison of spectra obtained on Dec. 9 and Dec. 15, 2009 allows us to roughly correct for the time delay of 7.8 days \citep{2011Courbinal} between these two images. We found that the differences decrease but do not vanish. Because spectra separated by the exact time delay are not available, we cannot conclude on the nature of the small spectral differences observed between $B$ and $C$ (Fig.~\ref{fig:he0435_median}$d,e$).

Apart from the spectral differences between the line profiles of some lensed images, the flux ratios do not vary with wavelength over the observed wavelength range, in agreement with the absence of significant differential extinction (at least over this small wavelength range).

\subsection{Microlensed part of the spectrum}

We used the macro-micro decomposition (MmD) method (\citealp{2007Sluseal}, \citealp{2010Hutsemekersal}, \citealp{2012Sluseal}) to separate the part of the quasar spectrum that is affected by microlensing from the part which is not. The MmD interprets the differences between the spectra of a pair of gravitationally lensed images, and in particular the differences between the line profiles, under the hypothesis that microlensing affects the continuum emission, but does not affect the emission line, or at least any wavelength interval of it (cf. appendix~A in \citealp{2010Hutsemekersal}). This method allows one to determine the macro-magnification ratio $M$ between the macro-images of the quasar and a microlensing factor $\mu$, which quantifies the additional (de-)magnification caused by microlensing. \vspace{1mm}

\begin{figure}
\resizebox{\hsize}{!}{\includegraphics[width=\textwidth]{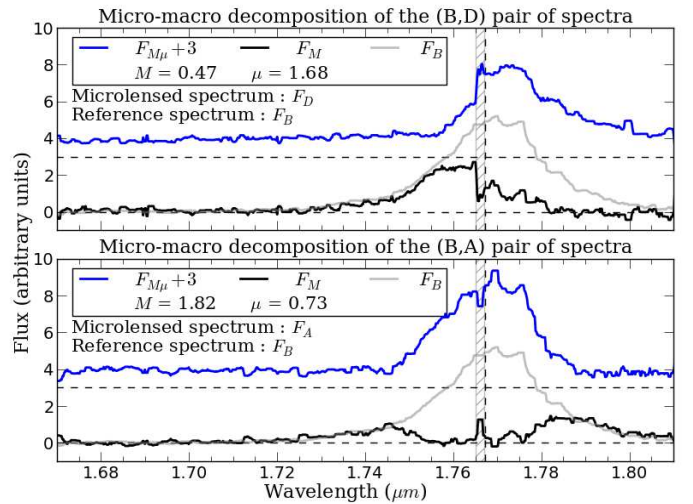}}
\caption{\textit{Upper panel:} decomposition of the quasar spectrum into a macrolensed-only component ($F_M$, black line) and a component both macro- and microlensed ($F_{M\mu}$, blue line) by applying MmD to the spectra of images $B$ and $D$. $F_{M\mu}$ is shifted upward by three units. The spectrum of image $B$ ($F_B$, grey line), assumed to be unaltered by microlensing, is shifted downward till its continuum intensity is null and is superimposed on $F_M$. The spectra are filtered with a five-pixel-wide median filter for clarity. The vertical dotted line indicates the redshifted rest wavelength of the \ce{H\alpha} line. \textit{Lower panel:} same for the pair of spectra of images $B$ and $A$.}
\label{fig:micro_macro}
\end{figure}

{$\bullet$ \raggedleft{}\textbf{Image \textit{D}: undergoing a large magnification}} \\
In Fig.~\ref{fig:micro_macro} (upper panel), we applied the MmD to the median spectra of images $B$ and $D$, computed using all the available spectra. We derived a macro-magnification ratio $M = 0.47 \pm 0.03$ and a microlensing factor $\mu = 1.68 \pm 0.10$.

Microlensing magnifies the continuum of image $D$  as well as the red wing of its \ce{H\alpha} line, which causes the displacement of its peak towards longer wavelengths (see Fig.~\ref{fig:he0435_median}$f$). Microlensing affects about $ 50\%$ of the flux of the Balmer emission line.

Since the time delay between $B$ and $D$ is 6.5 days \citep{2011Courbinal}, we used the spectra obtained on Dec. 9 and Dec. 15 to remove intrinsic variability. The results obtained following this procedure are noisier but consistent with the above analysis. This is expected because the time delay is shorter than the time scale of large-amplitude intrinsic variations. This supports the microlensing interpretation of the spectral differences observed in the \ce{H\alpha} emission line.

Microlensing has been detected in image $D$ by previous studies. \citet{2003Wisotzkial} found signatures of microlensing in $D$, using the first spatially resolved spectroscopic observations of HE0435-1223, which were secured in 2002. Applying the MmD to visible spectra acquired in 2004, \citet{2012Sluseal} detected microlensing of the \ce{CIII]} and \ce{MgII} lines in the spectrum of image $D$. \vspace{1mm}

{$\bullet$ \raggedleft{}\textbf{Image \textit{A}: weakly de-magnified}} \\
The MmD applied to the pair of spectra $A$-$B$ is displayed in Fig.~\ref{fig:micro_macro} (lower panel). The continuum emission and the core of the \ce{H\alpha} line are de-magnified by microlensing. The microlensing factor is $\mu = 0.73 \pm 0.05$. The macro-magnification ratio between $A$ and $B$ is $M=1.82 \pm 0.10$.

As in the previous section, we used spectra secured on Dec. 9 and Dec. 15 to remove intrinsic variability. But, since $A$ and $B$ are separated by an eight-day time delay \citep{2011Courbinal}, this is only an approximate correction. The MmD applied to these spectra leads to noisier but consistent results, supporting our microlensing interpretation. 

Interestingly, microlensing affects the line profile symmetrically in $A$, in contrast to the effect that alters the spectrum of $D$. About $60\% $ of the flux of the \ce{H\alpha} line is microlensed. This is not unexpected since de-magnification can act on more extended regions than magnification (e.g. \citealp{2004LewisIbata}).

That there is microlensing in image $A$ has been suggested based on photometric monitoring \citep{2011Riccial,2011Courbinal}. Using narrow-band photometric data obtained in 2007, \citet{2011Mosqueraal} have detected chromatic microlensing in the continuum of image $A$.

The comparison of the flux ratio $A/B= 1.43 \pm 0.03$ measured in the H band (Table~\ref{tab:lit_ratios}) with the contemporary R-band measurement $A/B = 1.53 \pm 0.05$ of \citet{2011Courbinal} suggests a weak chromatic trend, but we find no indication of chromaticity when the H-band $A/B$ flux ratio is compared with the V-, R- and I-band measurements made by \citet{2011Riccial} at a slightly different epoch (Aug.-Sep. 2009).

\section{Discussion and conclusions}
\label{sec:discussion}

\subsection{Comparison with previous studies}

The $A/B$ macro-magnification ratio we determined is consistent with the $L'$ flux ratio $A/B = 1.72 \pm 0.23$ measured by \citet{2011FadelyKeeton}. Our measurement confirms the $A/B$ flux ratio anomaly (e.g., \citealp{2012Fadely}).

The macro-magnification ratio $M_{D/B}= 0.47 \pm 0.03$ we derived is significantly lower than $M_{D/B}=0.73 \pm 0.015$ and $0.68 \pm 0.02$ obtained by \citet{2012Sluseal}, in the \ce{CIII]} and \ce{MgII} line profiles respectively, and than the $L'$ flux ratio $D/B=0.82\pm0.16$ measured by \citet{2011FadelyKeeton}.

Such a low $D/B$ macro-magnification ratio implies a micro-magnification of the continuum stronger than previously thought, both in the UV-visible and $L'$ band\footnote{In the $L'$ band, 20\% of the continuum may come from the accretion disk; therefore microlensing may not be negligible \citep{2013Sluseal}.}. It also implies that microlensing affects a larger part of the \ce{CIII]} and \ce{MgII} lines than found by \citet{2012Sluseal} using MmD. This method minimizes the part of the BLR emission that is microlensed and thus provides a lower limit of the microlensing factor. Such a low $M_{D/B}$ may not be easily reproduced by smooth mass models.

On the other hand, we also investigated microlensing scenarios corresponding to $M_{D/B} \sim 0.7$ and $M_{D/B} \sim 0.8$, obtained by \citet{2012Sluseal} and \citet{2011FadelyKeeton}. In brief, for $M_{D/B} \sim 0.8$, the micro-de-magnification of only the blue wing of the \ce{H\alpha} line would reproduce the line profile distortions observed in image $D$, while for $M_{D/B} \sim 0.7$, microlensing would affect the whole line profile, magnifying the \ce{H\alpha} red wing and simultaneously de-magnifying its blue wing (details in the appendix).

None of these scenarios appears completely satisfactory, given the available data but, regardless of the scenario considered, this does not change the strong microlensing dichotomy observed between the blue and red wings of \ce{H\alpha} in image $D$.

\begin{figure}
\includegraphics[width=0.5\textwidth , height=3.5cm]{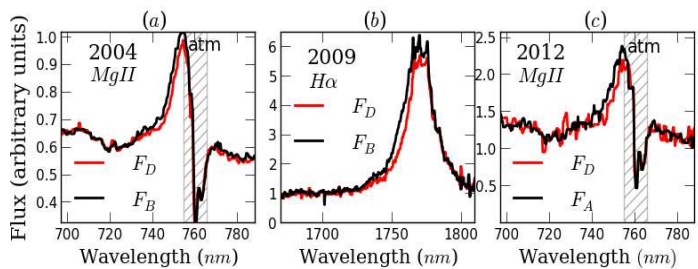}
\caption{($a$) Comparison of the \ce{MgII} line profiles in images $B$ and $D$, observed in 2004. ($b$) Comparison of the \ce{H\alpha} line profiles of images B and D, observed in 2009. ($c$) Comparison of the \ce{MgII} line profiles of images $A$ and $D$, observed in 2012. In every panel, the continua have been scaled so that they overlap. %The wavelength is in the observer rest frame.
}
\label{fig:comp_Dref_vis_ir}
\end{figure}

The microlensing effect causing the displacement of the peak in $D$ has already been observed by \citet{2012Sluseal} in the \ce{MgII} line profile, in 2004 (Fig.~\ref{fig:comp_Dref_vis_ir}$a$). It is still detected in the \ce{MgII} line profile obtained in 2012 with Magellan IMACS (Fig.~\ref{fig:comp_Dref_vis_ir}$c$)\footnote{Spectra of $A$ and $D$ were acquired simultaneously, separately from spectra of $B$ and $C$, with a different integration time (Table~\ref{tab:obsHE0435}). Thus, spectra of $B$ and $D$ cannot be safely compared with each other.}. This large-amplitude microlensing effect hence appears to be stable over time, in agreement with the microlensing time-scale estimated for the HE0435-1223 system: $\sim 30 \sqrt{M/M_{\astrosun}}$ years (considering a relative transverse velocity of $\unit{600}{\kilo\meter\per\second}$, \citealp{2010SchmidtWambganss}).

\subsection{Consequences for the BLR geometry}

A rotating disk is a popular model for Balmer-line-emitting region (e.g., \citealp{2005Smith}, \citealp{2009Bon}), but less flattened (e.g. \citealp{2009Gaskell}, \citealp{2012Goad}) and biconical (\citealp{2011Fischer}) geometries have also been proposed. The way microlensing alters the \ce{H\alpha} line in images $A$ and $D$ sets constraints on the BLR geometry and its velocity field.

Simulations \citep{1990SchneiderWambsganss} showed that only a non-spherically symmetric BLR can be at the origin of a blue/red dichotomous microlensing effect that causes the displacement of the peak, as observed in image $D$.

In image $D$, the most blueshifted and redshifted parts of the \ce{H\alpha} line, which correspond to the high-velocity emitting regions, are affected very differently by microlensing. Hence, the regions of the BLR that produce these parts of the line profile must be spatially well separated in projection. In addition, these highly Doppler-shifted parts of the line are not microlensed in image $A$, which implies that the corresponding emitting regions are not only distant from each other in projection, but also located away from the central continuum source. On the other hand, the core of the \ce{H\alpha} line, which corresponds to the low-velocity part of the BLR, is affected by microlensing in $A$, so that it most likely comes from a compact region close to the continuum source in projection.

A rotating-ring geometry for the \ce{H\alpha} emitting region fits these constraints nicely, favouring Keplerian-disk models for the low-ionization line-emitting region. Biconical winds cannot be completely ruled out, but we expect that only specific combinations of inclinations and velocity fields of the bicone will conform with both the type 1 nature of the system and the constraints derived from microlensing.

In future work, we will more quantitatively investigate the microlensing effects on flattened and biconical BLRs using simulated spectra. New spectra that simultaneously cover a wider wavelength range and contain several broad emission lines would also allow us to gather information on the ionization structure of the broad line region.

\tiny
\bibliographystyle{aa}
\bibliography{biblio}

\begin{thebibliography}{28}
\expandafter\ifx\csname natexlab\endcsname\relax\def\natexlab#1{#1}\fi

\bibitem[{{Abajas} {et~al.}(2002){Abajas}, {Mediavilla}, {Mu{\~n}oz},
  {Popovi{\'c}}, \& {Oscoz}}]{2002Abajasal}
{Abajas}, C., {Mediavilla}, E., {Mu{\~n}oz}, J.~A., {Popovi{\'c}}, L.~{\v C}.,
  \& {Oscoz}, A. 2002, \apj, 576, 640

\bibitem[{{Blackburne} {et~al.}(2011){Blackburne}, {Pooley}, {Rappaport}, \&
  {Schechter}}]{2011Blackburneal}
{Blackburne}, J.~A., {Pooley}, D., {Rappaport}, S., \& {Schechter}, P.~L. 2011,
  \apj, 729, 34

\bibitem[{{Bon} {et~al.}(2009){Bon}, {Popovi{\'c}}, {Gavrilovi{\'c}}, {Mura},
  \& {Mediavilla}}]{2009Bon}
{Bon}, E., {Popovi{\'c}}, L.~{\v C}., {Gavrilovi{\'c}}, N., {Mura}, G.~L., \&
  {Mediavilla}, E. 2009, \mnras, 400, 924

\bibitem[{{Courbin} {et~al.}(2011){Courbin}, {Chantry}, {Revaz}, {Sluse},
  {Faure}, {Tewes}, {Eulaers}, {Koleva}, {Asfandiyarov}, {Dye}, {Magain}, {van
  Winckel}, {Coles}, {Saha}, {Ibrahimov}, \& {Meylan}}]{2011Courbinal}
{Courbin}, F., {Chantry}, V., {Revaz}, Y., {et~al.} 2011, \aap, 536, A53

\bibitem[{{de Vaucouleurs}(1948)}]{1948DeVaucouleurs}
{de Vaucouleurs}, G. 1948, Annales d'Astrophysique, 11, 247

\bibitem[{{Eigenbrod} {et~al.}(2006){Eigenbrod}, {Courbin}, {Meylan},
  {Vuissoz}, \& {Magain}}]{2006Eigenbrod}
{Eigenbrod}, A., {Courbin}, F., {Meylan}, G., {Vuissoz}, C., \& {Magain}, P.
  2006, \aap, 451, 759

\bibitem[{{Fadely} \& {Keeton}(2011)}]{2011FadelyKeeton}
{Fadely}, R. \& {Keeton}, C.~R. 2011, \aj, 141, 101

\bibitem[{{Fadely} \& {Keeton}(2012)}]{2012Fadely}
{Fadely}, R. \& {Keeton}, C.~R. 2012, \mnras, 419, 936

\bibitem[{{Fischer} {et~al.}(2011){Fischer}, {Crenshaw}, {Kraemer}, {Schmitt},
  {Mushotsky}, \& {Dunn}}]{2011Fischer}
{Fischer}, T.~C., {Crenshaw}, D.~M., {Kraemer}, S.~B., {et~al.} 2011, \apj,
  727, 71

\bibitem[{{Gaskell}(2009)}]{2009Gaskell}
{Gaskell}, C.~M. 2009, \nar, 53, 140

\bibitem[{{Goad} {et~al.}(2012){Goad}, {Korista}, \& {Ruff}}]{2012Goad}
{Goad}, M.~R., {Korista}, K.~T., \& {Ruff}, A.~J. 2012, \mnras, 426, 3086

\bibitem[{{Hutsem{\'e}kers} {et~al.}(2010){Hutsem{\'e}kers}, {Borguet},
  {Sluse}, {Riaud}, \& {Anguita}}]{2010Hutsemekersal}
{Hutsem{\'e}kers}, D., {Borguet}, B., {Sluse}, D., {Riaud}, P., \& {Anguita},
  T. 2010, \aap, 519, A103

\bibitem[{{Kochanek} {et~al.}(2006){Kochanek}, {Morgan}, {Falco}, {McLeod},
  {Winn}, {Dembicky}, \& {Ketzeback}}]{2006Kochanekal}
{Kochanek}, C.~S., {Morgan}, N.~D., {Falco}, E.~E., {et~al.} 2006, \apj, 640,
  47

\bibitem[{{Lewis} \& {Ibata}(2004)}]{2004LewisIbata}
{Lewis}, G.~F. \& {Ibata}, R.~A. 2004, \mnras, 348, 24

\bibitem[{{Markwardt}(2009)}]{2009Markwardt}
{Markwardt}, C.~B. 2009, in Astronomical Society of the Pacific Conference
  Series, Vol. 411, Astronomical Data Analysis Software and Systems XVIII, ed.
  D.~A. {Bohlender}, D.~{Durand}, \& P.~{Dowler}, 251

\bibitem[{{Morgan} {et~al.}(2005){Morgan}, {Kochanek}, {Pevunova}, \&
  {Schechter}}]{2005Morganal}
{Morgan}, N.~D., {Kochanek}, C.~S., {Pevunova}, O., \& {Schechter}, P.~L. 2005,
  \aj, 129, 2531

\bibitem[{{Mosquera} {et~al.}(2011){Mosquera}, {Mu{\~n}oz}, {Mediavilla}, \&
  {Kochanek}}]{2011Mosqueraal}
{Mosquera}, A.~M., {Mu{\~n}oz}, J.~A., {Mediavilla}, E., \& {Kochanek}, C.~S.
  2011, \apj, 728, 145

\bibitem[{{Ricci} {et~al.}(2011){Ricci}, {Poels}, {Elyiv}, {Finet}, {Sprimont},
  {Anguita}, {Bozza}, {Browne}, {Burgdorf}, {Calchi Novati}, {Dominik},
  {Dreizler}, {Glitrup}, {Grundahl}, {Harps{\o}e}, {Hessman}, {Hinse},
  {Hornstrup}, {Hundertmark}, {J{\o}rgensen}, {Liebig}, {Maier}, {Mancini},
  {Masi}, {Mathiasen}, {Rahvar}, {Scarpetta}, {Skottfelt}, {Snodgrass},
  {Southworth}, {Teuber}, {Th{\"o}ne}, {Wambsgan{\ss}}, {Zimmer}, {Zub}, \&
  {Surdej}}]{2011Riccial}
{Ricci}, D., {Poels}, J., {Elyiv}, A., {et~al.} 2011, \aap, 528, A42

\bibitem[{{Richards} {et~al.}(2004){Richards}, {Keeton}, {Pindor}, {Hennawi},
  {Hall}, {Turner}, {Inada}, {Oguri}, {Ichikawa}, {Becker}, {Gregg}, {White},
  {Wyithe}, {Schneider}, {Johnston}, {Frieman}, \&
  {Brinkmann}}]{2004richardsal}
{Richards}, G.~T., {Keeton}, C.~R., {Pindor}, B., {et~al.} 2004, \apj, 610, 679

\bibitem[{{Schmidt} \& {Wambsganss}(2010)}]{2010SchmidtWambganss}
{Schmidt}, R.~W. \& {Wambsganss}, J. 2010, General Relativity and Gravitation,
  42, 2127

\bibitem[{{Schneider} \& {Wambsganss}(1990)}]{1990SchneiderWambsganss}
{Schneider}, P. \& {Wambsganss}, J. 1990, \aap, 237, 42

\bibitem[{{Sluse} {et~al.}(2007){Sluse}, {Claeskens}, {Hutsem{\'e}kers}, \&
  {Surdej}}]{2007Sluseal}
{Sluse}, D., {Claeskens}, J.-F., {Hutsem{\'e}kers}, D., \& {Surdej}, J. 2007,
  \aap, 468, 885

\bibitem[{{Sluse} {et~al.}(2012){Sluse}, {Hutsem{\'e}kers}, {Courbin},
  {Meylan}, \& {Wambsganss}}]{2012Sluseal}
{Sluse}, D., {Hutsem{\'e}kers}, D., {Courbin}, F., {Meylan}, G., \&
  {Wambsganss}, J. 2012, \aap, 544, A62

\bibitem[{{Sluse} {et~al.}(2013){Sluse}, {Kishimoto}, {Anguita}, {Wucknitz}, \&
  {Wambsganss}}]{2013Sluseal}
{Sluse}, D., {Kishimoto}, M., {Anguita}, T., {Wucknitz}, O., \& {Wambsganss},
  J. 2013, \aap, 553, A53

\bibitem[{{Smith} {et~al.}(2005){Smith}, {Robinson}, {Young}, {Axon}, \&
  {Corbett}}]{2005Smith}
{Smith}, J.~E., {Robinson}, A., {Young}, S., {Axon}, D.~J., \& {Corbett}, E.~A.
  2005, \mnras, 359, 846

\bibitem[{{van Dokkum}(2001)}]{2001Vandok_lacosmic}
{van Dokkum}, P.~G. 2001, \pasp, 113, 1420

\bibitem[{{Wisotzki} {et~al.}(2003){Wisotzki}, {Becker}, {Christensen},
  {Helms}, {Jahnke}, {Kelz}, {Roth}, \& {Sanchez}}]{2003Wisotzkial}
{Wisotzki}, L., {Becker}, T., {Christensen}, L., {et~al.} 2003, \aap, 408, 455

\bibitem[{{Wisotzki} {et~al.}(2001){Wisotzki}, {Christlieb}, {Liu}, {Maza},
  {Morgan}, \& {Schechter}}]{2001Wisotzkial}
{Wisotzki}, L., {Christlieb}, N., {Liu}, M.~C., {et~al.} 2001, in Astronomical
  Society of the Pacific Conference Series, Vol. 237, Gravitational Lensing:
  Recent Progress and Future Go, ed. T.~G. {Brainerd} \& C.~S. {Kochanek}, 63

\end{thebibliography}

\begin{acknowledgements}
TA acknowledges support from FONDECYT grant number 11130630. DS is supported by the German Deutsche Forschungsgemein-schaft, DFG project number SL172/1-1.
\end{acknowledgements}

\Online

\normalsize

\begin{table*}
\caption{Scientific observations of HE0435-1223.}
\label{tab:obsHE0435}
\begin{tabular}{lcccccc}
\hline \\ [-2.0ex]
Date & Images & Grating/Grism/Filter & Instument & Exp. time (s) & Seeing ('') & Airmass \\
 \hline  \\ [-2.0ex]
11/10/2004 & B,D & G300V+GG435 & FORS1 (ESO-VLT) & $4 \times 1400$ & 0.48 & 1.03 \\
11/11/2004 & B,D & G300V+GG435 & FORS1 (ESO-VLT) & $2 \times 1400$ & 0.57 & 1.11 \\
19/10/2009 & A,B,C,D & H & SINFONI (ESO-VLT) & $2 \times 600$ & 0.74 & 1.06 \\
19/10/2009 & A,B,D   & H & SINFONI (ESO-VLT) & $2 \times 600$ & 0.91 & 1.03 \\
09/12/2009 & A,B,C,D & H & SINFONI (ESO-VLT) & $5$\tablefootmark{a}$ \times 600$ & 0.48 & 1.11 \\
09/12/2009 & A,B,D   & H & SINFONI (ESO-VLT) & $4 \times 600$ & 0.49 & 1.20 \\
10/12/2009 & B,C & H & SINFONI (ESO-VLT) & $4 \times 600$ & 0.58 &  1.04 \\
15/12/2009 & A,B,C,D & H & SINFONI (ESO-VLT) & $2 \times 600$ & 0.73  & 1.15 \\
15/12/2009 & A,B,D & H & SINFONI (ESO-VLT) & $2 \times 600$ & 0.76  & 1.24 \\
20/09/2012 & A,D & Gra-300 lines/mm & IMACS (Magellan-Baade) & $2 \times 1800$ & 0.74 & 1.49 \\
20/09/2012 & B,C & Gra-300 lines/mm & IMACS (Magellan-Baade) & $1 \times 1800$ & 0.88 & 1.14 \\
20/09/2012 & B,C & Gra-300 lines/mm & IMACS (Magellan-Baade) & $1 \times 1200$ & 0.94 & 1.07 \\
\hline
\end{tabular}\newline
\tablefoottext{a}{One exposure has been discarded because of target misalignment.}
\end{table*}

\begin{table*}
\begin{center}
\caption{Flux ratios of the HE0435-1223 components measured at several epochs (compiled from the literature).}
\label{tab:lit_ratios}
\begin{tabular}{lcccccc}
\hline
Date & HJD\tablefootmark{a} range & Band & $A/B$ & $C/B$ & $D/B$ & References  \\ \hline
\rule{0pt}{3ex}2003 Aug. & 2870 & $V$ & $1.70 \pm 0.22$ & $0.93 \pm 0.08$ & $0.88 \pm 0.12$ & 1 \\
2003 Aug. & 2870  & $I$ & $1.67 \pm 0.11$ & $0.99 \pm 0.04$ & $0.81 \pm 0.06$ & 1 \\
2004 Jan. & 3013-3036 & $R$ & $1.73 \pm 0.05$ & $1.02 \pm 0.02$ & $0.88 \pm 0.02$ & 2 \\
2004 Jan. & 3015 & $H$ & $1.57 \pm 0.10$ & $1.00 \pm 0.03$ & $0.79 \pm 0.05$ & 1 \\
2007 Sep. & 4366 & $J$ & $1.58 \pm 0.09$ & $1.01 \pm 0.08$ & $0.84 \pm 0.06$ & 3 \\
2007 Sep. & 4366 & $H$ & $1.54 \pm 0.08$ & $1.03 \pm 0.08$ & $0.83 \pm 0.06$ & 3 \\
2007 Sep. & 4366 & $K$ & $1.42 \pm 0.03$ & $1.02 \pm 0.04$ & $0.79 \pm 0.03$ & 3 \\
2007 Oct. & 4316-4340 & $R$ & $2.16 \pm 0.04$ & $1.10 \pm 0.02$ & $0.91 \pm 0.02$ & 2 \\
2008 Jul.-Oct. & 4675-4744 & $V $& $1.58 \pm 0.08$ & $1.06 \pm 0.06$ & $0.85 \pm 0.05$ & 4 \\
2008 Jul.-Oct. & 4675-4744 & $R$ & $1.56 \pm 0.05$ & $1.06 \pm 0.05$ & $0.84 \pm 0.04$ & 4 \\
2008 Jul.-Oct. & 4674-4766 & $R$ & $1.66 \pm 0.04$ & $1.10 \pm 0.03$ & $0.86 \pm 0.02$ & 2 \\
2008 Jul.-Oct. & 4675-4744 & $I$ & $1.51 \pm 0.08$ & $1.06 \pm 0.06$ & $0.82 \pm 0.04$ & 4 \\
2008 Aug.-Dec. & 4682-4829 & $R$ & $1.63 \pm 0.05$ & $1.09 \pm 0.03$ & $0.86 \pm 0.02$ & 2 \\
2008 Aug. & 4709 & $K$  & $1.45 \pm 0.18$ & $0.79 \pm 0.04$ & $0.67 \pm 0.07$ & 5 \\
2008 Dec. & 4822 & $L'$ & $1.72 \pm 0.23$ & $1.01 \pm 0.08$ & $0.82 \pm 0.16$ & 5 \\
2009 Aug.-Sep. & 5064-5094 & $V$ & $1.41 \pm 0.04$ & $1.03 \pm 0.02$ & $0.82 \pm 0.02$ & 4 \\
2009 Aug.-Sep. & 5064-5094 & $R$ & $1.43 \pm 0.05$ & $1.05 \pm 0.03$ & $0.81 \pm 0.03$ & 4 \\
2009 Aug.-Sep. & 5047-5099 & $R$ & $1.53 \pm 0.03$ & $1.08 \pm 0.03$ & $0.82 \pm 0.02$ & 2 \\
2009 Aug.-Sep. & 5064-5094 & $I$ & $1.45 \pm 0.05$ & $1.05 \pm 0.03$ & $0.79 \pm 0.03$ & 4 \\
2009 Oct.-Dec. & 5111-5176 & $R$ & $1.53 \pm 0.05$ & $1.07 \pm 0.04$ & $0.80 \pm 0.03$ & 2 \\
2009 Oct.-Dec. & 5123,5174,5175,5180 & $H$ & $1.43 \pm 0.03$ & $1.02 \pm 0.03$ & $0.72 \pm 0.02$ & 6 \\
\hline 
\end{tabular}
\end{center}
\tablefoot{
\tablefoottext{a}{HJD = Julian Day - 2450000}
}
\tablebib{
(1) \citet{2006Kochanekal}; (2) \citet{2011Courbinal}; (3) \citet{2011Blackburneal}; (4) \citet{2011Riccial}; (5) \citet{2011FadelyKeeton}; (6) This work: the photometric flux ratios were computed by multiplying the spectrum of each quasar image by the transmission of the $H$-band filter and then integrating the flux.
}
\end{table*}

\FloatBarrier

\begin{appendix}

\renewcommand{\thesection}{}
\section{ Microlensing scenarios for different \textit{D}/\textit{B} macro-magnification ratios}
\label{app:microlensing_scenar}

We investigated microlensing scenarios corresponding to the macro-magnification ratios derived by \citet{2012Sluseal}, $M_{D/B} \sim 0.7$, and \citet{2011FadelyKeeton}, $M_{D/B} \sim 0.8$.

\begin{figure}
\includegraphics[width=0.5\textwidth]{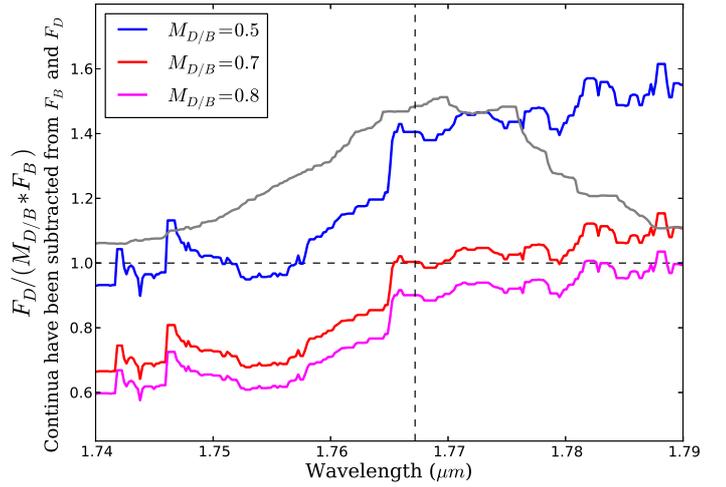}
\caption{Ratio between the \ce{H\alpha} line profiles observed in the spectrum of image D and the spectrum of image $B$, corrected for the macro-magnification ratio. The continua that compose the spectra of images $B$ and $D$ were previously subtracted. This ratio is a proxy for the effect of microlensing on the \ce{H\alpha} line profile in $D$. At a given wavelength, a ratio higher than unity corresponds to the micro-magnification of the BLR emission. Conversely, a lower ratio indicates a micro-de-magnification. The spectra were smoothed with a fifteen-pixel-wide median filter for clarity.}
\label{fig:proxy_miro_Ms}
\end{figure}

For $M_{D/B} \sim 0.7$ (and $\mu \sim 1.1$ in the continuum), the whole line profile would be affected by microlensing so that a rigorous MmD cannot be performed. Simultaneous micro-magnification of the \ce{H\alpha} red wing (although to a lesser extent than for $M_{D/B} \sim 0.5$) and micro-de-magnification of the blue wing would reproduce the distortions observed in the \ce{H\alpha} line profile of image $D$ (red curve in Fig. \ref{fig:proxy_miro_Ms}). Caustics can cause such microlensing effect since they can delineate magnification and de-magnification areas.

For $M_{D/B} \sim 0.8$, only the blue wing of \ce{H\alpha} is micro-de-magnified (magenta curve in Fig. \ref{fig:proxy_miro_Ms}) and the continuum is not microlensed  ($\mu \sim 1$ in the continuum). A similar microlensing effect that magnifies only the broad emission line region, has presumably been observed for the system J1004-4142 \citep{2004richardsal}. In that interpretation, a de-magnification area stands close enough to affect the BLR but far enough not to affect the continuum source. This agrees with the stability of the $D/B$ flux ratio over time and wavelength (see Table~\ref{tab:lit_ratios}). However, considering the unification model for active galactic nuclei, in which a dusty torus surrounds the BLR (regardless of its geometry), we would expect a broad de-magnification region to de-magnify the dusty torus as well. Thus, the $L'$-band flux should be affected, which is not observed.

\newpage 

\end{appendix}

\end{document}